\documentclass{emulateapj}
\usepackage{apjfonts}

\newcommand{\bdv}[1]{\mbox{\boldmath$#1$}}
\newcommand{\bd}[1]{{\rm #1}}
\def\rel{{\rm rel}}
\def\e{{\rm E}}
\def\au{{\rm AU}}
\def\muas{{\mu\rm as}}
\def\kms{{\rm km}\,{\rm s}^{-1}}
\def\kpc{{\rm kpc}}
\def\ss{\scriptstyle}
\begin{document}
\title{OGLE-2003-BLG-262: Finite-Source Effects from a Point-Mass
Lens \altaffilmark{*}}

\author{Jaiyul Yoo\altaffilmark{1},
D.L. DePoy\altaffilmark{1},
A. Gal-Yam\altaffilmark{2},
B.S. Gaudi\altaffilmark{3},
A. Gould\altaffilmark{1},
C. Han\altaffilmark{4},
Y. Lipkin\altaffilmark{2},
D. Maoz\altaffilmark{2},
E.O. Ofek\altaffilmark{2},
B.-G. Park\altaffilmark{5},
and R.W. Pogge\altaffilmark{1} \\
(The $\mu$FUN Collaboration) \\ and \\
A.~Udalski\altaffilmark{6},
I.~Soszy{\'n}ski\altaffilmark{6},
{\L}.~Wyrzykowski\altaffilmark{6},
M.~Kubiak\altaffilmark{6},
M.~Szyma{\'n}ski\altaffilmark{6},
G.~Pietrzy{\'n}ski\altaffilmark{6},
O.~Szewczyk\altaffilmark{6},
and K.~\.Zebru{\'n}\altaffilmark{6} \\
(The OGLE Collaboration)}
\altaffiltext{1}
{Department of Astronomy, The Ohio State University,
140 West 18th Avenue, Columbus, OH 43210; jaiyul, depoy, gould, 
pogge@astronomy.ohio-state.edu}
\altaffiltext{2}
{School of Physics and Astronomy and Wise Observatory, Tel Aviv University,
Tel Aviv 69978, Israel; avishay, yiftah, dani, eran@wise.tau.ac.il}
\altaffiltext{3}
{Harvard-Smithsonian Center for Astrophysics, Cambridge, MA 02138; 
sgaudi@cfa.harvard.edu}
\altaffiltext{4}
{Department of Physics, Institute for Basic Science Research,
Chungbuk National University, Chongju 361-763, Korea;
cheongho@astroph\-.chungbuk.ac.kr}
\altaffiltext{5}
{Korea Astronomy Observatory,
61-1, Whaam-Dong, Youseong-Gu, Daejeon 305-348, Korea; bgpark@boao.re.kr}
\altaffiltext{6}
{Warsaw University Observatory, Al.~Ujazdowskie~4, 00-478~Warszawa, Poland;
udalski, soszynsk, wyrzykow, mk, msz, pietrzyn, szewczyk, 
zebrun@astrouw.edu.pl}
\altaffiltext{*}
{Based in part on observations obtained with the 1.3~m Warsaw
Telescope at the Las Campanas Observatory of the Carnegie Institution
of Washington.}

\slugcomment{accepted for publication in the Astrophysical Journal}

\begin{abstract}
We analyze OGLE-2003-BLG-262,  a relatively short, $t_\e=12.5\pm0.1$ day,
microlensing event generated
by a point-mass lens transiting the face of a K giant source in the Galactic
bulge.  We use the resulting finite-source effects to measure the
angular Einstein radius, $\theta_\e=195\pm17\,\muas$, and so constrain the
lens mass to the full-width half-maximum interval 
$0.08 < M/M_\odot < 0.54$.
The lens-source relative proper motion
is $\mu_\rel = 27\pm2\,\kms\,\kpc^{-1}$.  Both values are typical of what
is expected for lenses detected toward the bulge.  Despite the short duration
of the event, we detect marginal evidence for a ``parallax asymmetry'',
but argue that this is more likely to be induced by acceleration of
the source, a binary lens, or possibly
by statistical fluctuations.  
Although OGLE-2003-BLG-262 is only the second published event to date
in which the lens transits the source, such events will
become more common with the new OGLE-III survey in place.  We therefore
give a detailed account of the analysis of this event to facilitate
the study of future events of this type.
\end{abstract}


\section{Introduction}
\label{intro}

Immediately following the announcement of the first microlensing
detections \citep{macho93,eros93,ogle93}, three groups independently showed
that one could measure the microlens angular Einstein radius, 
\begin{equation}
\theta_\e=\sqrt{\kappa M\pi_\rel},\qquad 
\kappa\equiv {4G\over c^2\,\au}\simeq 8\,{{\rm mas}\over M_\odot},
\label{eqn:thetaedef}
\end{equation}
from the deviations on the microlensing lightcurve
induced by the finite size of the source \citep{gould94,nw,wm}.
Here $M$ is the mass of the lens and $\pi_\rel$ is the lens-source
relative parallax.
Although all three considered the case of a point-mass lens passing
close to or over the face of the source star, the great 
majority of the actual $\theta_\e$ measurements made over the ensuing
decade used binary-lens events in which the source passed over
the binary caustic 
\citep{albrow99,albrow00,albrow01,afonso00,alcock00,jin}.
There has been only one single-lens event for which finite-source
effects have yielded a measurement of $\theta_\e$.  This was the spectacular
event MACHO-95-30, whose M4~III source of radius $r_*\sim60r_\odot$
was transited
by the lens \citep{alcock97}.  In fact, of the more than 1000 single-lens 
microlensing events
discovered to date, only two have a measured $\theta_\e$ by any technique.
The other was the equally spectacular MACHO-LMC-5
whose source-lens relative proper motion $\mu_\rel$ was measured by directly
imaging and resolving the source and the
M-dwarf lens six years after the event \citep{alcock01}.  The angular Einstein
radius was then inferred from,
\begin{equation}
\theta_\e = \mu_\rel~t_\e, 
\label{eqn:tedef}
\end{equation}
where $t_\e$ is the Einstein crossing time, 
which had been measured during the event.

Measurements of $\theta_\e$ are important because they constrain
the physical properties of the lens.
For most events, the only measured parameter that is related to the
physical properties of the lens is $t_\e$, which 
(from eqs.~[\ref{eqn:thetaedef}] and [\ref{eqn:tedef}]) is a combination
of three such properties, $M$, $\pi_\rel$, and $\mu_\rel$.  If 
$\theta_\e$ is measured, one then determines $\mu_\rel$ from 
equation~(\ref{eqn:tedef}), and the only
remaining ambiguity is between $M$ and $\pi_\rel$ 
(see eq.~[\ref{eqn:thetaedef}]).  In some cases, $\mu_\rel$ is 
directly of interest.
For example, measurement of the proper motion of
the binary event 
MACHO-98-SMC-1 led to the conclusion that the lens was in the SMC
itself rather than the Galactic halo \citep{afonso00}.  

In other cases, one can combine the measurement of $\theta_\e$
with other measurements or limits to further constrain the character
of the lens.  The most dramatic example of this would be measurement
of the microlens parallax,
\begin{equation}
\pi_\e = \sqrt{\pi_\rel\over\kappa M},
\label{eqn:piedef}
\end{equation}
which can be determined either by observing the event from a satellite
in solar orbit \citep{refsdal66,gould95} or from the distortion of
the microlens lightcurve induced by the accelerated motion of the Earth
\citep{gould92}.  If both $\theta_\e$ and $\pi_\e$ are measured,
one completely solves the event. That is,
\begin{equation}
M = {\theta_\e^2\over \kappa\pi_\rel},\qquad \pi_\rel=\theta_\e\pi_\e.
\label{eqn:mandpirel}
\end{equation}
Unfortunately, while $\pi_\e$ has been measured for about a dozen
events, only one of these also has a firm measurement of $\theta_\e$
\citep{jin}, although \citet{smw} also obtained tentative measurements of both
$\theta_\e$ and $\pi_\e$.

	Another type of constraint that can be combined with a measurement
of $\theta_\e$ is an upper limit on the lens flux, which can often be
obtained from the lightcurve.   This flux limit can be converted into a 
luminosity limit at each possible lens distance.  If the lens is assumed 
to be a main-sequence star, then using equation (\ref{eqn:thetaedef})
and some reasonable assumption about the source distance, one can put
an upper limit on the lens mass  (e.g., \citealt{albrow00b}).   
Even in the absence of any additional
constraints, equation~(\ref{eqn:thetaedef}) can be combined with
a Galactic model to make statistical statements about the lens properties
(e.g., \citealt{alcock97}).

	The principal reason that most $\theta_\e$ measurements come
from binary lenses and that single-lens measurements are extremely rare
is that the ratio $\rho$, of the angular source radius, $\theta_*$, to
the Einstein radius,
\begin{equation}
\rho \equiv {\theta_*\over\theta_\e},
\label{eqn:rhodef}
\end{equation}
is usually 
extremely small.  At the distance of the Galactic bulge, even a clump
giant has an angular radius $\theta_*\sim 6\,\muas$, and main-sequence stars
are an order of magnitude smaller.  By contrast, typical Einstein radii
are $\theta_\e \sim 310\,\muas[(M/0.3M_\odot)(\pi_\rel/40\,\muas)]^{1/2}$.
Hence, the probability that the lens will pass directly over the source,
which is what is required for substantial finite source effects
\citep{gouldwelch}, is very small.  By contrast, binary lenses, with
their extended caustic structures, have a much higher probability of
generating finite-source effects.

	However, new microlensing surveys are beginning to alter this
situation.  In particular, the new phase of the 
Optical Gravitational Lens Experiment, OGLE-III \citep{ogleIII}, with
its 
dedicated 1.3~m telescope and new $35'\times 35'$ field, $0.\hskip-2pt''26$
pixel, mosaic CCD camera
and generally excellent image quality is generating microlensing alerts
at the rate of 500/season (as reported by the OGLE-III Early Warning
System, EWS, http://www.astrouw.edu.pl/\-$\sim$ogle/ogle3/ews/ews.html),
roughly an order of magnitude higher than previous surveys.  
This high event rate is itself enough to overcome the low,
${\cal O}(\rho)$, probability of a source-crossing event, and so to generate
a few finite-source affected EWS alerts per year.  Moreover, because
EWS relies on image-subtraction \citep{wozniak}, it is sensitive
to extremely high magnification events of relatively faint sources,
which have a higher chance of a source crossing than do typical events.

OGLE-III is able to generate this high event rate only by reducing
its visits to individual fields below 1/night.  Hence, it would
not customarily observe the alerted event during the lens transit of the
source.
However, several groups, including the Probing Lensing Anomalies NETwork
(PLANET, \citealt{planet}), the Microlensing Planet Search 
(MPS, \citealt{mps}), and the Microlensing Follow-Up Network 
($\mu$FUN, http://www.astronomy.ohio-state.edu/$\sim$microfun/)
intensively monitor alerts from EWS and also from the 
Microlensing Observations in
Astrophysics collaboration (MOA, \citealt{moa}), primarily to search for
planets.  High-magnification events are the most sensitive to planetary
perturbations \citep{gouldloeb,gs}, so these groups tend to focus on
these events, particularly their peaks.  As a consequence, there
is a good chance they will detect finite-source effects when they occur.
Moreover, OGLE-III diverts time from its regular field rotation
(survey mode) to especially interesting events (follow-up mode)
and so can itself also directly
detect these effects.

Here we report observations of EWS alert OGLE-2003-BLG-262, which exhibited
clear indications of finite-source effects near its peak on 2003 July 19.
By fitting this event to a single-lens finite-source model, we measure the
$\theta_\e$ and so $\mu_\rel$. We use this information, combined with
a measurement of $t_\e$ to constrain the mass of the lens. We also present
marginal ($\ga 3\,\sigma$) evidence for an asymmetry which, if due to parallax
effects, would imply that the lens was a brown dwarf. However, we argue that
the observed asymmetry is either due to statistical fluctuations, a weak
binary lens, or acceleration of the source. Our analysis provides a framework
in which to analyze future finite-source single-lens event, which should
be considerably more common due to the higher rate of alerted events.

\section{Observational Data}
\label{data}
The microlensing event OGLE-2003-BLG-262 was identified by the OGLE-III
EWS \citep{udal} on  2003 June 26, i.e., more than three weeks  before
peak, which occurred on HJD$' \equiv$ HJD$-2450000 = 2839.84$ over the
Pacific Ocean.  OGLE-III observations were carried out with the  1.3-m
Warsaw telescope at the Las Campanas Observatory, Chile, which is
operated by the Carnegie Institution of Washington.  These comprise a
total of 170 observations in $I$ band, including 68 in the 2001 and 2002
seasons.  The exposures were generally the standard  120 s, except for
three special 40 s exposures on the peak and following night when the
star was too bright for the standard exposure time. Photometry was obtained
with the OGLE-III image subtraction technique data pipeline 
\citep{ogleIII} based in part on the \citet{wozniak} DIA implementation.
The source had also been monitored by OGLE-II and was found to be very
stable over  four previous seasons (April~1997 -- October~2000).

Following the alert, the event was monitored by $\mu$FUN
from sites in Chile and Israel.  The Chile observations were carried out at
the 1.3m (ex-2MASS) telescope at Cerro Tololo InterAmerican Observatory,
using the ANDICAM, which simultaneously images at optical
and infrared wavelengths \citep{depoy}.  
During the seven nights from HJD$'$ 2838.5 to 2844.8,
there were a total of 45 observations in $I$, 4 in $V$, and 28 in $H$.
The $I$ and $V$ observations were generally 300 s, although the  
exposures were shortened to 120 s during the three nights from 
2839.6 to 2841.8.
The individual $H$ observations were 60 s and were grouped in 5 dithered
exposures, which were taken simultaneously with one 300 s $V$ or $I$ exposure
or with two 120 s $I$ exposures.  All images were flat fielded using sky flats
for $V$ and $I$, and dome flats for $H$.
Photometry was  obtained with DoPHOT \citep{dophot} for all $V$, $I$, and $H$
images.

After reductions, the contiguous groups of 5 (or 10 in the case of 
back-to-back $V$ and $I$ exposures) $H$ points were averaged 
into single data points to yield the above-stated 28 points.

The $\mu$FUN Israel observations were carried out on the Wise 1m telescope
at Mitzpe Ramon, 200 km south of Tel-Aviv, roughly $105^\circ$
east of Chile.  During the nights of 
$\sim 2839.3$, $\sim 2841.3$, and $\sim 2842.3$, there were a total
of four observations in $I$ and three in $V$.  The exposures (all 240 s) were
obtained using the Wise Tektronix 1K CCD camera.  Data were flat-fielded
and zero corrected in the usual way, and photometry obtained with DoPHOT.

\begin{figure}[t]
\centerline{\epsfxsize=3.5truein\epsffile{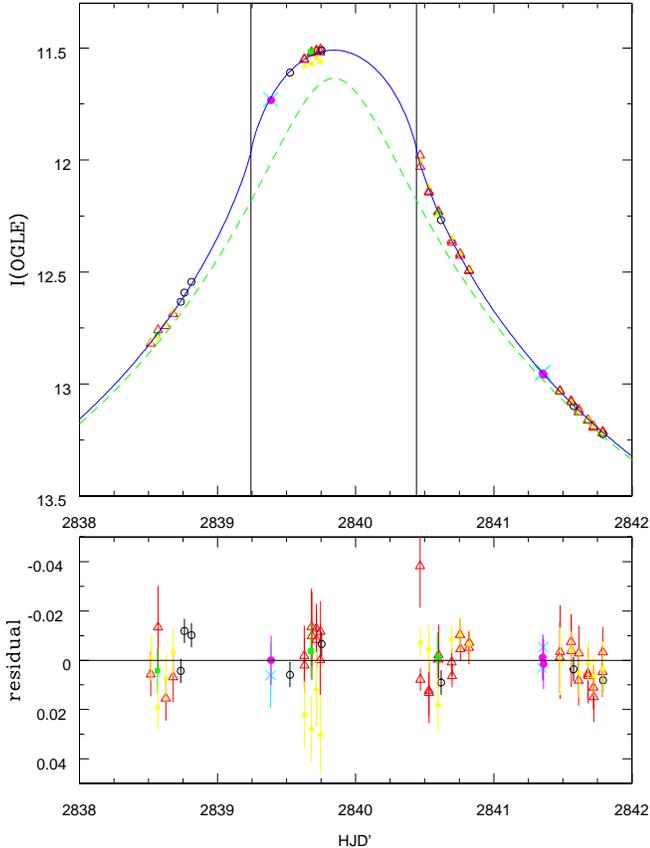}}
\caption{\label{fig:lightcurve1}
Photometry of microlensing event OGLE-2003-BLG-262 near its
peak on 2003 July 19.34 (HJD 2452839.83).  Data points are in $I$
(OGLE: empty circles; $\mu$FUN Chile: empty triangles; $\mu$FUN Israel: 
crosses),
$V$ ($\mu$FUN Chile: filled squares; $\mu$FUN Israel: filled circles), and $H$
($\mu$FUN Chile: filled triangles).  All bands are linearly rescaled so
that $F_s$ and $F_b$ are the same as the OGLE observations, which
define the magnitude scale.  When the lens is close to or inside
(vertical lines) the source, the lightcurves are expected to
differ due to limb darkening (LD).  The solid curve shows the best fit
model for the $I$-band curve.  The fact that the $H$-band points
near the peak are below this curve is in qualitative accord with the lower
LD in $H$ compared to $I$.  The dashed curve shows the lightcurve
expected for the same lens model, but a point source.}
\end{figure}

The position of the source is $\rm R.A.=17^h57^m08.\!\!^s51$,
$\rm decl.=-30^\circ20'05.\!\!\arcsec1~(J2000)$
$(l,b = 0.41918,-3.46935)$, and so was accessible
for most of the nights near peak from Chile, but only a few hours
from Israel.  Unfortunately, due to a communications mixup, $\mu$FUN 
Chile observations on the peak night were bunched in a narrow time
interval.  Happily, when these are combined with the two OGLE observations
and the one $I$-band  and one $V$-band $\mu$FUN Israel observations, 
the rising half of the
peak is still clearly traced out.  See Figure~\ref{fig:lightcurve1}.

\begin{figure}[b]
\centerline{\epsfxsize=3.5truein\epsffile{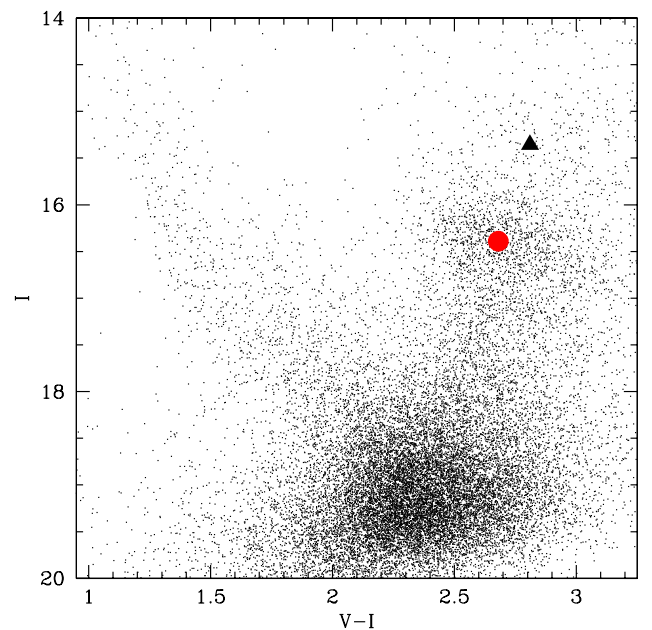}}
\caption{\label{fig:cmd}
Calibrated color-magnitude diagram (CMD) of a $10'$ square around
OGLE-2003-BLG-262 taken from OGLE-II photometry well before the event.
The source (marked with a black triangle) is about 1 mag brighter and
0.2 mag redder than the centroid of the clump giants (marked with a red
circle).  The fit shows
negligible blending, so the apparent source position on the CMD is
virtually identical to its true (deblended) position.}
\end{figure}

We emphasize that while three of the datasets have relatively 
few points,  two of these small datasets actually play 
crucial roles.  The three post-peak $\mu$FUN Israel $I$ points
serve to align this dataset with the two larger $I$ datasets and so
enable the first point (on 2839.38) to directly test the near-peak
finite-source profile, which otherwise would be determined by
a single compact set of points.  See Figure~\ref{fig:lightcurve1}.
The four $\mu$FUN Chile $V$ points allow determination of the color of the 
source  and so permit one to estimate the source size and thus the proper
motion and angular Einstein radius.  See \S~\ref{model}.  With
only three points, two of which are nearly coincident, the $\mu$FUN Israel 
$V$ data do not contribute significantly to the fit because they are absorbed
by two fitting parameters, $F_s$ and $F_b$.  However, they are included here
for completeness.

The source lies in one of the OGLE-II calibrated photometry 
fields \citep{ogleIImaps} 
(ftp://bulge.princeton.edu/\-ogle/ogle2/maps/bulge/) and
this allows us to place it on a calibrated color magnitude diagram.
See Figure~\ref{fig:cmd}.  The source lies on the red giant branch, 
about 1 mag brighter than the clump and about 0.2 mag redder.
It therefore has considerably larger angular radius than typical microlens
sources, and this, together with the high magnification of the event,
considerably increased the chance for significant finite source
effects. 

\citet{sumi} measured the proper motion of the source (relative to the frame
of the Galactic bulge) and found $(\mu_{\alpha,s},\mu_{\delta,s})$ 
$=(0.45\pm0.41,-5.75\pm0.40)~{\rm mas\ yr^{-1}}$. When corrected to the
Tycho-2 frame, this becomes 
$(\mu_{\alpha,s},\mu_{\delta,s})$ $=(-2.9,-12.4)~{\rm mas\ yr^{-1}}$.
 
\section{Formalism}
\label{form}

\subsection{Finite-Source Effects}
\label{sec:finite}
In most cases, the lensed star is regarded as a point source because the
angular size of the source is negligibly small compared to the angular 
separation of the source and the lens. The magnification is then given by
\citep{bod},
\begin{equation}
A(u)={u^2+2\over u(u^2+4)^{1/2}},
\label{eqn:amp}
\end{equation}
where $u$ is the projected source-lens separation in units of the angular
Einstein radius $\theta_\e$. However, this approximation breaks
down for $u\la \rho$. Finite-source effects then dominate.

\begin{figure}[t]
\centerline{\epsfxsize=3.5truein\epsffile{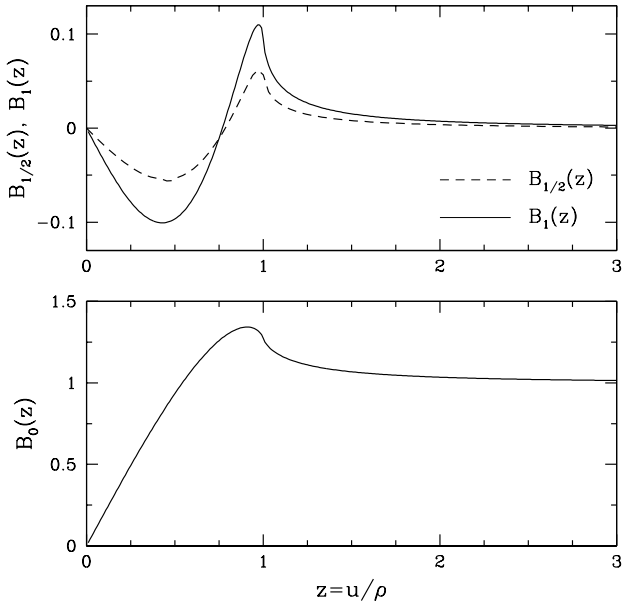}}
\label{fig:limblaw}
\caption{Finite source functions $B_0(z)$, $B_{1/2}(z)$ and $B_1(z)$ given by
eqs.~(\ref{eqn:b0eval}), (\ref{eqn:b12def}) and (\ref{eqn:b1def}).  
For $\rho\ll 1$,
the limb-darkened magnification is very well represented by
$A_\bd{ld}(u|\rho) = A(u)[B_0(z) - \Gamma B_1(z)]$, where
$\rho$ is the source size and $u$ is the lens-source separation, both in
units of $\theta_\e$, $\Gamma$ is the linear limb-darkening coefficient,
and $z=u/\rho$.}
\end{figure}

If the source were of uniform brightness, the total magnification would 
simply be the mean magnification over the source,
\begin{equation}
A_\bd{uni}(u|\rho) = W_0\left[(u/\rho)|\rho;A(x)\right],
\label{eqn:aunieval}
\end{equation}
where
\begin{equation}
W_n\left[z|\rho;f(x)\right] \equiv {1\over \pi}\int_0^{2\pi} \!\!\!\!\!\!\!\!
\bd{d}\theta \int_0^1 \!\!\!\!\!\!
\bd{d}r~r (1 - r^2)^{n/2} f(\rho\sqrt{r^2 + z^2 - 2 r z \cos\theta}).
\label{eqn:wndef}
\end{equation}
\citet{wm} gave an exact evaluation of this
expression in closed, albeit cumbersome, form. \citet{gould94}
advocated a simple approximation to equation (\ref{eqn:wndef}),
\begin{equation}
A_\bd{uni}(u|\rho) \simeq A(u)B_0(u/\rho), 
\qquad B_0(z)\equiv {z\rho}W_0\left[z|\rho;x^{-1}\right],
\label{eqn:uniapprox}
\end{equation}
which follows from the fact that $A(u)\simeq u^{-1}$ when $u\ll1$.
Note that $B_0$ depends on $\rho$ only through the ratio $z=u/\rho$.
However, \citet{gould94} did not explicitly evaluate $B_0$ nor did he
demonstrate the range of validity of the approximation~(\ref{eqn:uniapprox}).
It is straightforward to show that,
\begin{equation}
B_0(z) = {4\over\pi} z E(k,z),\qquad k\equiv {\rm min}(z^{-1},1),
\label{eqn:b0eval}
\end{equation}
where $E$ is the incomplete elliptic integral of the second kind, and where
we follow the notation of \citet{gr}.  Using the expansion
$A(u) = u^{-1}[1 + (3/8)u^2 + \ldots]$, and 
after some algebra, one may show that to second order in $\rho$,
\begin{equation}
A_\bd{uni}(u|\rho) = A(u)B_0(z)\biggl[1 + {\rho^2\over 8}Q(z)\biggr], \qquad
z\equiv {u\over \rho},
\label{eqn:auni2nd}
\end{equation}
where,
\begin{equation}
Q(z) = {1\over 3}\biggl[7 - 8z^2 - 4(1-z^2){F(k,z)\over E(k,z)}\biggr],
\label{eqn:qzdef}
\end{equation}
and where $F$ is the incomplete elliptic integral of the first kind.
We find numerically that $-0.38\leq Q(z)\leq 1$, where the limits are
saturated at $z=0.97$ and $z=0$ respectively.  For OGLE-2003-BLG-262, 
$\rho^2/8\la 5\times 10^{-4}$, which is about an order of magnitude 
smaller than our photometric errors.  The zeroth-order approximation
(\ref{eqn:uniapprox}) is therefore appropriate here and, we believe,
is likely to be appropriate in most other cases as well.

\subsection{Limb Darkening}
\label{sec:limb}

Real stars are not uniform, but rather are limb-darkened.
For simplicity and also because the quality of the data does not warrant
a more sophisticated treatment, we adopt a one-parameter 
linear limb-darkening law for the surface brightness of the source,
\begin{equation}
S_\lambda(\vartheta)=\bar S_\lambda\left[1-\Gamma_\lambda\left(1-{3\over2}
\bd{cos}~\vartheta\right)\right],
\label{eqn:limb}
\end{equation}
where $\vartheta$ is the angle between the normal to the stellar surface
and the line of sight, $\bar S_\lambda$ is the mean surface brightness
of the source, and the $\Gamma_\lambda$ is the limb-darkening (LD) coefficient
for a given wavelength band $\lambda$.
The factor 3/2 originates from our requirement that the
total flux be $F_{\bd{tot},\lambda}=\pi\theta_*^2\bar S_\lambda$.

The LD magnification is then (exactly),
\begin{eqnarray}
A_\bd{lld}(u|\rho)&=&W_0\left[(u/\rho)|\rho;A(x)\right] \\
&-&\Gamma \left\{ W_0\left[(u/\rho)|\rho;A(x)\right] - \nonumber
1.5\,W_1\left[(u/\rho)|\rho;A(x)\right] \right\}.
\label{eqn:aldexact}
\end{eqnarray}
However, we adopt the same simplifying approximation as above and write,
\begin{equation}
A_\bd{lld}(u|\rho)\simeq A(u)\left[B_0(z) - \Gamma\,B_1(z)\right],\qquad
\label{eqn:aldapprox}
\end{equation}
where
\begin{equation}
B_1(z) = B_0(z) - {3\over 2}{z\rho}W_1\left[z|\rho;x^{-1}\right].
\label{eqn:b1def}
\end{equation}

Figure~\ref{fig:limblaw} shows $B_0$, $B_{1/2}$ (see below) and $B_1$ 
as functions of $z$.
Note that $B_0(z)\to1$ and $B_1(z)\to0$ in the limit $z\to\infty$ so that
the magnification (eq.~[\ref{eqn:aldapprox}]) reduces to the point-source
case. In the opposite limit, $z\to0$, equations~(\ref{eqn:b0eval})
and (\ref{eqn:b1def}) reduce to
$B_0(z)\to2z$ and $B_1(z)\to(2-3\pi/4)z$, so that 
$A_\bd{fin}(0)=2/\rho\left[1+(3\pi/8-1)\Gamma\right]$. Hence, the peak
magnification depends primarily on $\rho$ and only weakly on $\Gamma$.

In high-precision LD measurements, it is generally accepted that
a two-parameter square-root LD law is more appropriate to describe brightness
profiles of stars \citep{albrow99,dale} than equation~(\ref{eqn:limb})
although it is not used in the present work. Therefore, for completeness
we extend the above formalism to a
two-parameter square-root LD law in the form of
\begin{equation}
S_\lambda(\vartheta)=\bar S_\lambda\left[1-\Gamma_\lambda\left(1-{3\over2}
\bd{cos}~\vartheta\right)-\Lambda_\lambda\left(1-{5\over4}
\bd{cos}^{1/2}~\vartheta\right)\right],
\end{equation}
where $\Lambda_\lambda$ is the additional LD coefficient for a given
wavelength band $\lambda$. The magnification can then be approximated by,
\begin{equation}
A_\bd{sqrtld}(u|\rho)\simeq A(u)\left[B_0(z)-\Gamma\,B_1(z)-\Lambda\,B_{1/2}(z)
\right],
\end{equation}
where
\begin{equation}
B_{1/2}(z)=B_0(z)-{5\over4}z\rho W_{1/2}\left[z|\rho;x^{-1}\right].\
\label{eqn:b12def}
\end{equation}

\subsection{Parallax Effects}
\label{sec:paral}

\begin{deluxetable*}{lrcrrcrrc}
\tablewidth{0pt}
\tablecaption{OGLE-2003-BLG-262 Fit Parameters\label{tab:par}}
\tablehead{\colhead{} & \multicolumn{2}{c}{Free Fit} & \colhead{} &
\multicolumn{2}{c}{Fixed LD} & \colhead{} &
\multicolumn{2}{c}{Fixed LD \& $\pi_\bd{E}$} \\ 
\cline{2-3} \cline{5-6} \cline{8-9} \\
\colhead{Parameter} & \colhead{Value} & \colhead{Error} & \colhead{} &
\colhead{Value} & \colhead{Error} & \colhead{} &
\colhead{Value} & \colhead{Error} }
\startdata
$t_0$(days)........ & 2839.8411 & 0.0015 
& & 2839.8415 & 0.0015 & & 2839.8424 & 0.0014\\
$u_0$................. & 0.0365 & 0.0005 
& & 0.0362 & 0.0004 & & 0.0360 & 0.0004\\
$t_\bd{E}$(days)........ & 12.5309 & 0.0945 
& & 12.5568 & 0.0941 & & 12.6181 & 0.0916\\
$\rho$................... & 0.0605 & 0.0010 & & 0.0599 & 0.0005 
& & 0.0595 & 0.0005\\
$\Gamma_{V}$................ & 0.8515 & 0.2069 & & 0.7200 
& - & & 0.7200 & -\\
$\Gamma_{I}$................. & 0.6118 & 0.1499 & & 0.4400 
& - & & 0.4400 & -\\
$\Gamma_{H}$................ & 0.0975 & 0.2028 & & 0.2600 
& - & & 0.2600 & -\\
$\pi_{\bd{E},\parallel}$.............. & $-$0.8572 & 0.3130 & & $-$0.8335 
& 0.3120 & & 0.0000 & -\\
$(F_b/F_s)_{I_1}$...... & $-$0.0011 & 0.0095 & & 0.0028 & 0.0093 &
& 0.0083 & 0.0091\\ 
$(F_b/F_s)_{I_2}$...... & $-$0.0275 & 0.0175 & & $-$0.0134 
& 0.0172 & & $-$0.0027 & 0.0168\\
$(F_b/F_s)_{I_3}$...... & 0.1865 & 0.0783 & & 0.2283 & 0.0746 &
& 0.2361 & 0.0749\\
$(F_b/F_s)_{V_2}$..... & 0.0122 & 0.0481 & & 0.0192 & 0.0478 &
& 0.0296 & 0.0479\\
$(F_b/F_s)_{V_3}$..... & 0.0406 & 0.1688 & & 0.1251 & 0.1465 &
& 0.1191 & 0.1473\\
$(F_b/F_s)_{H}$...... & $-$0.0048 & 0.0176 & & $-$0.0114 
& 0.0175 & & $-$0.0009 & 0.0172\\
$\chi^2$................. & 233.50 & - & & 252.66 & - & & 259.76 & -\\
\enddata

\tablecomments{Observatories: 1=OGLE, 2=$\mu$FUN Chile, 3=$\mu$FUN Israel}

\end{deluxetable*}

Microlensing events are fit to
\begin{equation}
F(t) = F_s A[u(t)] + F_b,
\label{eqn:foft}
\end{equation}
where $F_s$ is the source flux, $F_b$ is the blended background light, and
\begin{equation}
u(t) = \sqrt{[\tau(t)]^2 + [\beta(t)]^2}.
\label{eqn:uoft}
\end{equation}
Conventionally, rectilinear motion is assumed,
\begin{equation}
\tau(t) = {t - t_0\over t_\e},\qquad \beta(t) = u_0.
\label{eqn:taubeta}
\end{equation}
Hence, the simplest fit has five parameters, $F_s$, $F_b$, the
impact parameter $u_0$, the time of closest approach $t_0$, and
the Einstein timescale $t_\e$.  However, even if the source and
lens are in rectilinear motion, the Earth is not.  Thus, strictly 
speaking one should write
\begin{equation}
\tau(t) = {t - t_0\over t_\e} + \pi_{\e,\parallel} a_\parallel(t),
+ \pi_{\e,\perp} a_\perp(t),
\label{eqn:taupar}
\end{equation}
\begin{equation}
\beta(t) = u_0 - \pi_{\e,\parallel} a_\perp(t)
+ \pi_{\e,\perp} a_\parallel(t).
\label{eqn:betapar}
\end{equation}
Here $\bdv{a}\equiv (a_\parallel,a_\perp)$ is the difference in the
Earth's position (projected onto the plane of the sky and measured
in AU) relative to what it would have been had the Earth maintained
the velocity it had had at $t_0$, and the $a_\parallel$ direction is
defined by the direction of the Earth's (projected) acceleration
at $t_0$.

Choosing the Earth frame at the peak of the event as the inertial
frame is certainly not standard procedure.
It is more common, and mathematically
more convenient, to use the Sun's frame.  However, for 
relatively short events $t_\e \la {\rm yr}/2\pi$, the parallax effect
is quite weak, and it is only possible to measure one component of
$\bdv{\pi}_\e = (\pi_{\e,\parallel},\pi_{\e,\perp})$, namely the parallax 
asymmetry, which is the component ($\pi_{\e,\parallel}$) of the parallax
parallel to the Earth's projected acceleration at the peak of the
event \citep{gmb}.  In this case, $u_0$, $t_0$, and $t_\e$ as seen from
the Earth at the event peak are very well defined by the fit to the
event without parallax, whereas these quantities as seen from the
Sun are impossible to determine.  For these short events, $a_\perp\sim 0$,
and the impact of $a_\parallel$ through $\beta$ is undetectable because it is
absorbed into $u_0$, $t_\e$, $F_s$, and $F_b$.
Equations~(\ref{eqn:taupar})
and (\ref{eqn:betapar}) then reduce to,
\begin{equation}
\tau(t) = {t - t_0\over t_\e} + \pi_{\e,\parallel} a_\parallel(t),
\qquad \beta(t) = u_0.
\label{eqn:taubetapar}
\end{equation}

\section{Model Fitting}
\label{model}

We begin by fitting the event taking account of both LD and 
parallax. There are then a total of
20 free parameters: 12 parameters for $F_s$ and $F_b$ from
each of the six observatory/filter combinations, 3 LD parameters,
one each for $I$, $V$ and $H$, the basic microlensing parameters $t_0$, $u_0$,
and $t_\e$, as well as the source size, $\rho$, and the parallel component
of the parallax, $\pi_{\e,\parallel}$.  We consider the
possibility of a correction for seeing, but find no correlation of the
residuals of the $\mu$FUN Chile $I$ or $H$ data with seeing.  The source
is quite bright (see Fig.~\ref{fig:cmd}) and it is virtually
unblended (see below), so it is quite plausible that there would be
no seeing correlations.  We set a minimum error of 0.003 magnitudes
for all observations, regardless of what value the photometry programs
report.  We then rescale the errors for the OGLE, $\mu$FUN Chile $I$,
and $H$ by factors of 1.62, 1.12, and 1.83 respectively, in order
to force $\chi^2/$dof to unity.  There are too few points in each of
the remaining observatory/filter combinations to permit accurate
rescalings, and the actual total $\chi^2$ values for these are consistent
with the reported errors being correct.  

We minimize $\chi^2$ using Newton's method 
(e.g., \citealt{numrec}), which guarantees that one has found a
local minimum because the derivative of $\chi^2$ with respect
to each parameter is zero. In contrast to 
caustic-crossing binary lenses \citep{albrow99b,dom}, and to space-based 
\citep{gould94b,refsdal66} and ground-based \citep{smp}
parallax measurements for which there can be multiple
local minima, standard microlensing (even when modified by inclusion
of finite-source effects) is expected to have a single global minimum.
We nevertheless checked for multiple minima by adopting several initial
trial solutions that were consistent with point-source/point-lens
fits to a data set that excluded the peak.  All converged to the same solution.

\begin{figure}[t]
\centerline{\epsfxsize=3.5truein\epsffile{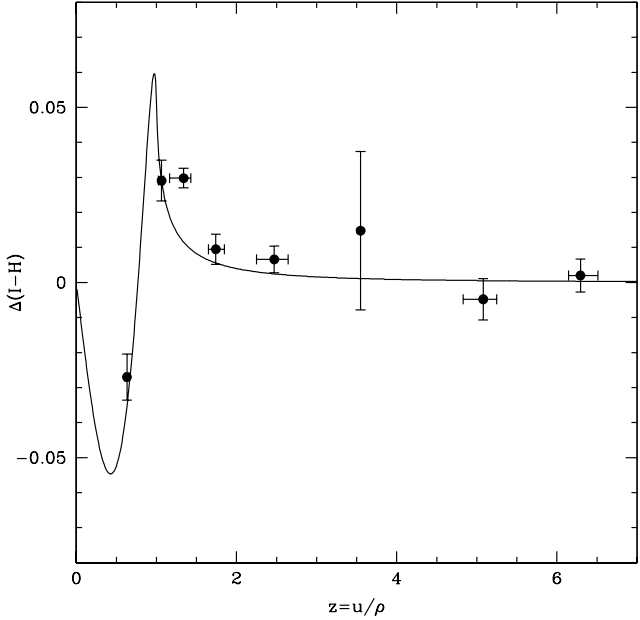}}
\caption{\label{fig:ld}
Model-independent color changes due to limb-darkening.
A linear regression of $H$ on $I$ flux is performed at high $z$
($z>1.7$) to put the two passbands on the same scale and to remove
the small blending difference.  Then $I-H$ is measured at each point
and the measurements for each day are averaged, except for HJD$'\sim 2840.5$,
($z\sim 1.25$), which is broken into two bins.  The curve is
$0.5 B_1(z)$, which is the expected form of this magnitude difference for
a linear limb-darkening difference $\Gamma_I-\Gamma_H=0.5$.}
\end{figure}

We initially allow the three LD coefficients to be
free parameters.  We find fit values  and errors
$(\Gamma_V,\Gamma_I,\Gamma_H) = (0.85\pm 0.21,0.61\pm 0.15,0.10\pm 0.20)$
(see Tab.~\ref{tab:par}).
These errors are all relatively large. The values therefore appear only
mildly inconsistent with those of EROS-BLG-2000-5,
$(\Gamma_V,\Gamma_I,\Gamma_H) = (0.72,0.44,0.26)$ \citep{dale},
a slightly redder source with much better measured LD.  It is then
somewhat shocking to discover that there is
a net penalty of $\Delta\chi^2=19$ for enforcing the EROS-BLG-2000-5 LD values.
A major part of the problem is that while the errors in the individual
LD parameters are large, the data strongly demand a large LD 
{\it difference}
$\Delta \Gamma =\Gamma_I -\Gamma_H = 0.51 \pm 0.09$ when $\Gamma_I$ is 
held fixed at 0.44. 
That is, although the errors on the individual determinations of 
$\Gamma_I$ and $\Gamma_H$ are large, they are strongly correlated, such
that the difference $\Delta\Gamma$ is much better determined.
This in turn can be traced to the fact that there is a color offset 
$\Delta (I-H) = -0.03$ at the peak, which is clearly visible in 
Figure~\ref{fig:lightcurve1} and which the fitting routine ascribes
to the source having much more LD in $I$ than $H$ and hence being relatively
blue in the center.  See Figure~\ref{fig:ld}.   
However, since the measurement seems to contradict what is otherwise known
about LD, and derives primarily from a single cluster of data points, which
may be subject to  common systematic error, we choose to fix the three
$\Gamma$'s at the above stated EROS-BLG-2000-5 values.
We thereby lose any independent LD information.  This is not
a major loss since our errors are too large to be competitive with
other measurements (e.g. \citealt{dale}).  Our main concern is that
whatever problem may be corrupting the LD could also impact the measurement
of the parameters that we are most interested in measuring, which are
principally $\rho$ and $t_\e$.  In fact, by enforcing these $\Gamma$'s,
$\rho$ is changed by only 1.6\% and $t_\e$ by only 0.3\%.  Since 
enforcing the LD parameters has no practical consequences (other than
the loss of LD information), we adopt the EROS-BLG-2000-5 value.  

\begin{figure}[b]
\centerline{\epsfxsize=3.5truein\epsffile{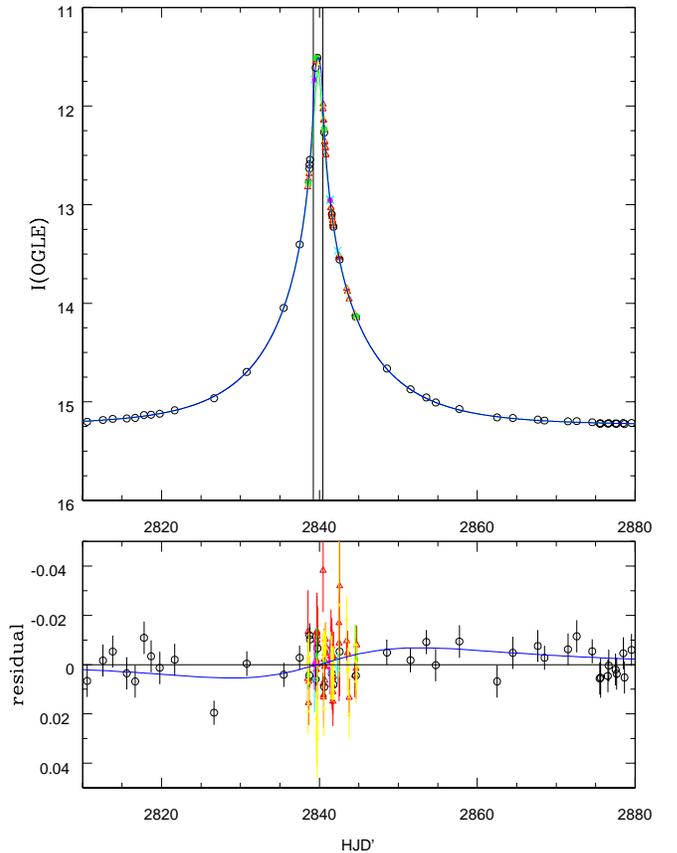}}
\caption{\label{fig:lightcurve2}
Similar to Fig.~\ref{fig:lightcurve1}, but now a
full view of the lightcurve of OGLE-2003-BLG-262 over about 5 Einstein
timescales.  The fit
does not include parallax, and the residuals (lower panel) show an asymmetry
such as would be induced by acceleration of the Earth parallel
to the direction of lens motion (solid curve).}
\end{figure}

We then find,
\begin{equation}
\rho= 0.0599\pm 0.0005,\qquad t_\e = 12.557 \pm 0.094\, {\rm days}.
\label{eqn:rhoteeval}
\end{equation}
Figure \ref{fig:lightcurve1} shows the fit to the data in the region
of the peak.  All six observatory/band combinations have been linearly
rescaled to have an $F_s$ and $F_b$ equal to those of the OGLE data set.  
The three $I$ band data
streams should then follow the same lightcurve, whose best fit
model is shown by the solid curve.  However, the $V$ and $H$ band
points should deviate from this curve during the source 
crossing,  $|t - t_0| \la \rho~t_e = 0.76\,$days, because of LD.
There are not enough data in the $V$ band to test this.
As mentioned above, the $H$ band cluster of points near the peak 
clearly lies below the curve, probably by too much.

We measure a parallax asymmetry,
\begin{equation}
\pi_{\e,\parallel}= 0.83\pm 0.31.
\label{eqn:pieeval}
\end{equation}
That is, parallax is formally detected at the $3\,\sigma$ level.
More specifically, $\Delta\chi^2=7$ 
relative to enforcing $\pi_{\e,\parallel}=0$.
To illustrate the strength (or lack thereof) of this detection we
show in Figure \ref{fig:lightcurve2} the fit to the data enforcing
$\pi_{\e,\parallel}=0$.  The lower panel of this figure shows the residuals
together with their expected form for $\pi_{\e,\parallel}=1$.  When
we first constructed this figure on about HJD$'=2870$, we realized that
there might still be time to test the reality of this parallax detection.
OGLE observations were then intensified from one every several days to
one or two per day.  These additional observations did not tend to
confirm the detection, but also did not firmly contradict it.  Hence,
the parallax detection remains ambiguous.  All previous events with
firm parallax detections had Einstein timescales at least 5 times longer
than this one, so it would have been remarkable if we had obtained
a robust detection. Moreover, as we discuss in \S~\ref{sec:par},
the sign of the effect is opposite to what would be produced by the expected
lens-source kinematics, while the effect itself could be produced
by xallarap or by lens binarity.

The errors shown in Table~\ref{tab:par} are $\sqrt{c_{ii}}$ where
$c_{ij}$ is the $ij$-th element of covariant matrix, and 
the correlation coefficients 
defined as $\tilde{c}_{ij}\equiv c_{ij}/\sqrt{c_{ii}}\sqrt{c_{jj}}$ are,
\begin{equation}
\left(\begin{array}{rrrrrrr}
\ss 1.0000 &\ss 0.2000&\ss-0.0675&\ss-0.1441&\ss 0.2288&\ss 0.1139&\ss-0.1137\\
\ss 0.2000 &\ss 1.0000&\ss-0.8463&\ss 0.7969&\ss-0.1861&\ss 0.8972&\ss-0.8971\\
\ss-0.0675 &\ss-0.8463&\ss 1.0000&\ss-0.9244&\ss 0.2790&\ss-0.9793&\ss 0.9758\\
\ss-0.1441 &\ss 0.7969&\ss-0.9244&\ss 1.0000&\ss-0.2990&\ss 0.9086&\ss-0.9060\\
\ss 0.2288 &\ss-0.1861&\ss 0.2790&\ss-0.2990&\ss 1.0000&\ss-0.2467&\ss 0.2531\\
\ss 0.1139 &\ss 0.8972&\ss-0.9793&\ss 0.9086&\ss-0.2467&\ss 1.0000&\ss-0.9986\\
\ss-0.1137 &\ss-0.8971&\ss 0.9758&\ss-0.9060&\ss 0.2531&\ss-0.9986&\ss 1.0000\\
\end{array}\right),
\label{corr}
\end{equation}
where parameters are $t_0$, $u_0$, $t_\e$, $\rho$, $\pi_{\e,\parallel}$, 
$(F_s)_{I_1}$, and $(F_b)_{I_1}$. As expected from experience with standard microlensing events,
$F_s$ and $F_b$ are extremely correlated, and these are both highly
correlated with $u_0$ and $t_\e$.  What is new in equation (\ref{corr})
is that $\rho$ is also highly correlated with these other four parameters.
The fundamental reason for this is that all five of these parameters are
symmetric in $(t-t_0)$.  By contrast $\pi_{\rm E,\parallel}$ is only
weakly correlated with the other parameters.

\section{Constraints on the Event}
\label{sec:constraints}

\subsection{Angular Einstein Radius $\theta_\e$}
\label{sec:thetae}

As discussed by \citet{albrow00}, one can determine $\theta_*$ from the
source's dereddened color and magnitude $[(V-I)_0,I_0]_s$ by first transforming
from $(V-I)_0$ to $(V-K)_0$ using the color-color relations of
\citet{BB88}, and then applying the empirical relation between color
and surface brightness to obtain $\theta_*$ \citep{vanb}.

Again following \citet{albrow00}, we determine $[(V-I)_0,I_0]_s$ from the
measured offset of the unamplified source (as determined from the
microlensing fit) relative to the centroid of the clump giants
on an instrumental CMD, the latter's
dereddened color and magnitude being regarded as ``known''.
We measure this offset to be
\begin{equation}
\Delta I = I_s - I_\bd{clump} = -1.06,~~~
\Delta (V-I) = (V-I)_s - (V-I)_\bd{clump} = 0.15.
\label{eqn:delclump}
\end{equation}
In general, the source may be blended, so that the $V$ and $I$ of the
source derived from the microlensing fit will not necessarily agree
with those of the object identified as the ``source'' in an image
taken at baseline.  Hence, one cannot in general derive the offset
from a CMD constructed from such a baseline image.  However, in this
case, there is essentially no blending, so the offset in the
baseline calibrated CMD shown in Figure~\ref{fig:cmd} is virtually 
identical (within 0.02 mag) to that given in equation~(\ref{eqn:delclump}).

The ``known'' values of $[(V-I)_0,I_0]_\bd{clump}$ have recently come
under dispute.  The basic problem is that the previous calibrations
of these quantities relied on a number of steps, in each of which
it was assumed that the ratio of total to selective extinction was
$R_{VI} \equiv A_V/E(V-I)\sim 2.5$.  However, using this same value,
\citet{pac98} and \citet{stutz} found respectively that the colors of bulge
clump giants and RR Lyrae stars were anomalous relative to local populations.
\citet{popow00} then proposed that these anomalies could be resolved
if the dust toward this line of sight were itself anomalous, with
$R_{VI}\sim 2.1$.  \citet{udal03} then demonstrated that this was
very likely the case based on OGLE-II data.  While it would be both
worthwhile and feasible
to retrace all the steps that led to the old calibration in light
of this revised $R_{VI}$, the magnitude of this project lies well beyond
the scope of the present work.  Pending such a revision, we adopt a
simpler approach.  

The distance to the Galactic center has now been measured geometrically
by \citet{eisen} to be $R_0 = 8.0\pm 0.4$ kpc based on the ``visual-binary''
method of \citet{sg}.  Bulge stars are of similar metallicity to
local stars, so the clump should be of similar color to the 
{\it Hipparcos}
clump stars $(V-I)_0\sim 1.00$.  (Recall that it was the apparent failure of 
this expectation that led to the discovery of anomalous extinction.)\ \
The $I$-band luminosity of clump stars does not depend strongly on
age (until the stars are so young that the turnoff luminosity approaches
that of the horizontal branch).  Hence, the bulge clump stars should
have approximately the same $M_I$ as the {\it Hipparcos} sample.
For this we adopt $M_I=-0.20$, the value found by \citet{pacstan} for
their 70 pc sample (and prior to their reddening correction which we
consider to be substantially too large.)  Hence, in lieu of a more
thoroughgoing calibration, we adopt
\begin{equation}
[(V-I)_0,I_0]_\bd{clump} = (1.00,14.32).
\label{eqn:viclump}
\end{equation}
Combining equations~(\ref{eqn:delclump}) and (\ref{eqn:viclump}) and
applying the \citet{vanb} relation,
we find
\begin{equation}
\theta_* = 11.7\pm 1.0\,\muas,
\label{eqn:thetastar}
\end{equation}
where the error comes primarily from the 8.7\% intrinsic scatter in
the \citet{vanb} relation.

\begin{figure}[t]
\centerline{\epsfxsize=3.5truein\epsffile{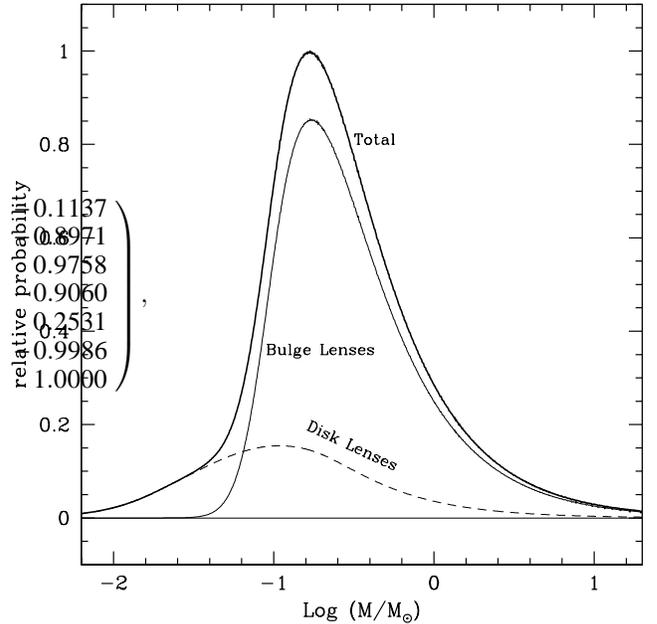}}
\caption{Constraints on the lens mass of OGLE-2003-BLG-262.  The
curves show the relative probability of different lens masses
given the measurement of $\theta_\e = 195\,\muas$ and the
mass distribution along the line of sight as predicted by the
\citet{hangould1,hangould2} model.  The solid and dashed curves 
show the probability for bulge-bulge and disk-bulge combinations
of lenses and sources.  The bold curve is their sum.
The constraints arising from the determination of 
$\mu_\rel$ (equivalently $t_\e$)
would be extremely weak and are not incorporated here.}
\label{fig:massdist}
\end{figure}

This evaluation would appear to depend on the assumption that
the source suffers exactly as much extinction as a typical clump star,
which it might not, either due to highly variable extinction or to
the source lying well in the foreground and so in front of a large
fraction of the dust.  In fact, if it were determined that the extinction
toward the source were greater than to the clump by $\Delta E(V-I) = 0.2$
or less by $\Delta E(V-I)=-0.6$, the estimate of $\theta_*$ would
change less than 3\%.  This is because the changes in the inferred
surface brightness and luminosity lead to changes in the source-size
estimate that go in opposite directions.

Combining equations~(\ref{eqn:rhoteeval}) and (\ref{eqn:thetastar}),
we obtain,
\begin{eqnarray}
\label{eqn:muteeval}
&&\theta_\e = 195\pm 17\, \muas, \\
&&\mu_\rel = 5.63 \pm 0.49\, \bd{mas}\,\bd{yr}^{-1} = 26.7 \pm 2.3\,
\kms\,\bd{kpc}^{-1}. \nonumber
\end{eqnarray}
We now use  this measurement of $\theta_\e$,
in conjunction with its definition, equation~(\ref{eqn:thetaedef}), to write
the source-lens relative parallax as a function of the lens mass,
\begin{equation}
\pi_\rel(M) = {\theta_\e^2\over\kappa M}=
4.8\,\muas\biggl({M\over M_\odot}\biggr)^{-1}.
\label{eqn:pirelofm}
\end{equation}
Given a Galactic mass model along the line of sight, $\rho(x)$, the 
prior probability of a given relative parallax is proportional to
\begin{equation}
P(\pi_\rel)\propto \int_0^\infty\!\!\!\!\!\!\bd{d} D_s\,D_s^2 \rho(D_s)
\int_0^{D_s}\!\!\!\!\!\!\bd{d} D_l\,D_l\,\rho(D_l)
\delta\biggl(\pi_\rel - \biggl[{\au\over D_l}-{\au\over D_s}\biggr]\biggr),
\label{eqn:pofpirel}
\end{equation}
where $D_l$ and $D_s$ are the lens and source distances.   We adopt
the \citet{hangould1,hangould2} model, and in 
Figure~\ref{fig:massdist} we plot $\pi_\rel(M) P[\pi_\rel(M)]$ versus $\log M$
to display the constraint placed on the mass by the measurement of
$\theta_\e$.  The full-width half-maximum range is,
\begin{equation}
\log (M/M_\odot) = -0.7\pm 0.4\qquad {\rm (FWHM)}.
\label{eqn:fwhm}
\end{equation}
In the absence of such a measurement, the only constraint
comes from the measurement of $t_\e$, and this is extremely weak, having
a full width half maximum of a factor $\sim 100$.  See figure 1
from \citet{gould00}.  Indeed, as shown in that figure, the mere
supposition that the lens is a star places stronger constraints on the
lens mass than does the measurement of $t_\e$.  

The measurements of $\theta_\e$ and $t_\e$ yield $\mu_\rel$ 
(eq.~[\ref{eqn:tedef}]).
Since the distribution of $\mu_\rel$ varies with $D_l$ and $D_s$, one
could in principle use its determination (eq.~[\ref{eqn:muteeval}]) to
place further constraints on combinations of these parameters and so
(through eq.~[\ref{eqn:pirelofm}])
on the mass.  In practice, for bulge sources, the distribution of
$\mu_\rel$ hardly varies as a function of lens position, even
when one considers bulge versus disk lenses.  Moreover, the actual
measured value of $\mu_\rel$ 
is near the peak of that distribution. Hence, we do
not incorporate this constraint.

The $\mu_\rel$ measurement does effectively rule out a foreground
disk source. (Without this constraint, i.e. from the CMD alone, the source
could plausibly be a disk clump giant at $D_s\sim 5\,\kpc$).  
However, for disk-disk
events along this line of sight, the observer, lens, and source all
share the same transverse motion due to the flat rotation curve of the
Galaxy.  Hence, only their peculiar motions relative to this rotation
enter $\mu_\rel$, and these are only of order 10's of $\kms$.   Hence,
$\mu_\rel$ would be only a few $\kms\,\kpc^{-1}$, much slower than
the measured value.

The \citet{sumi} proper-motion measurement of the source independently
rules out a foreground disk lens, since the source is moving roughly
opposite to the direction of Galactic rotation at about
$\mu\sim-v_c/R_0$, where $v_c\sim220~\kms$ and $R_0=8~\kpc$. In fact, this
measurement by itself would be consistent with the source lying in the 
background disk, behind the bulge. However, such a scenario is virtually
ruled out by the CMD (see Fig.~\ref{fig:cmd}), which shows the source lying
in or slightly above the bulge giant branch. If the source lay at, say,
10~kpc, it would intrinsically be $\sim$0.5 magnitude brighter still.

Combining our measurement $\mu_\rel=5.6~{\rm mas\ yr^{-1}}$, with 
the \citet{sumi}
measurement $(\mu_\alpha,\mu_\delta)$ $=(-2.9,-12.3)~{\rm mas\ yr^{-1}}$,
we can effectively rule out a disk lens. These measurements imply
$|\bdv{\mu}_L|=|\bdv{\mu}_s+\bdv{\mu}_\rel|\gtrsim7~{\rm mas\ yr^{-1}}$,
whereas a disk lens would be expected to have roughly zero proper motion.

\subsection{Lens Luminosity $M_{I,l}$}
\label{sec:lenslum}

The measurement of the unlensed background flux, $F_b$, gives an upper
limit to the flux from the lens.  The measured background flux is a
function of observatory and filter, and tends to grow with larger mean
seeing.  Hence, the best constraint is expected to come from the 
observatory with the best seeing.  In our case, this is OGLE.
The OGLE $F_b$ is also by far the best constrained, in part because
of the large number of baseline points.  The OGLE background-to-source
flux ratio is $F_b/F_s = 0.003\pm 0.009$, which yields a $3\,\sigma$
lower limit on the magnitude difference of the lens and source,
$I_l - I_s > 3.6$.  For this limit to be at all relevant, the
lens must be close to the turnoff or brighter, implying that it
is close to a solar mass.  Then, from equation~(\ref{eqn:pirelofm}),
the source and lens must be nearly the same distance.  This implies
in turn that the above limit on apparent-magnitude difference translates
directly into a limit on absolute-magnitude difference.  Since the source
is about 1 mag brighter than the clump, the constraint yields only
$M_{I,l}>2.4$, which is of very limited value.

\section{Microlens Parallax $\pi_E$}
\label{sec:par}
The detection of microlens parallax is marginal. We therefore begin
by investigating whether its tentatively detected value is consistent
with what else is known about the lens.  Given this orientation, and
for simplicity of exposition, we ignore the very large error in the
measurement.  Only one component of the parallax is measured.  We
therefore actually have a limit, not a measurement, of 
$\pi_\e\geq |\pi_{\e,\parallel}| = 0.86$.  Together with 
equations~(\ref{eqn:mandpirel}) and (\ref{eqn:muteeval}), this implies
$M\leq 0.03\,M_\odot$ and $\pi_\rel\geq 170~\muas$.  The source distance
cannot be much more than $D_s\sim 10~\kpc$,  partly because of the
low density of stars at greater distances and partly because it would
lie in an unpopulated portion of the CMD.  Hence, 
$\pi_l = \pi_\rel + \pi_s > 270~\muas$, i.e. $D_l< 3.7~\kpc$.
That is, the lens would be a disk brown dwarf.

Apart from the small peculiar velocity of each, the lens and Earth
are both rotating about the Galactic center at the same speed.
Hence, the lens should be seen moving against the bulge at
about $\sim 220\,\kms$ towards Cygnus, which is to say at a position
angle roughly $30^\circ$ east of north.  Because the dispersion of
bulge stars is about $90\,\kms$, this should also be approximately
the direction of lens motion relative to the source.

Since only one component of $\bdv{\pi}_\e$ is measured, all we can test is
the sign of this prediction.  From the post-peak residuals to the
fit without parallax (Fig.~\ref{fig:lightcurve2}), the Earth is accelerating
in the direction of the lens motion (thus slowing down the end of the
event).  On July 19 (roughly one month after opposition), this
is basically opposite the direction of the Earth's motion, and so
is basically toward the west.  Since the field is south of the
ecliptic, there is also a small component of this (projected) acceleration 
toward the south.  Hence, the position angle of the projected 
acceleration vector is about $260^\circ$, which is misaligned with
the expected direction of the lens motion by about $130^\circ$.
That is, the expected sign of $\pi_{\e,\parallel}$ is opposite to
what is expected.

While it remains possible that the peculiar velocities of the
lens and source conspire to produce this result, the statistical
significance of the parallax measurement is not high enough to warrant
its acceptance in the face of this strong contrary expectation.

Moreover, there are at least two other possible explanations for
this asymmetry apart from statistical fluctuations.  The
first is xallarap, distortions in the light curve due to accelerated
motion of the source rather than the lens.  Indeed, \citet{smp}
showed that any parallax effect could be mimicked by the orbital
motion of the source around an unseen companion.  When both
components of $\bdv{\pi}_\e$ are well measured, this possibility
can be largely discounted because the probability of the source
being in a binary with the same inclination, phase, and period
as the Earth's orbit is extremely small.  However, in the present
case, in which all that is detected is a single component
of acceleration, there is a very wide class of source binaries
that could mimic the observed parallax signal. Moreover, by
the arguments given in \S~\ref{sec:lenslum}, any source companion
on the main-sequence would be undetectable in the lightcurve (other
than through its effect accelerating the source).  The xallarap
hypothesis could be checked by radial-velocity measurements.

Still another possible source of the asymmetry is a very weak binary
lens.  \citet{gaudi} detected a similarly weak asymmetry in 
OGLE-1999-BLG-36 and were able to model this either with parallax
or with a low-mass companion to the lens.  Thus, 
asymmetric residuals can be attributed to several effects including
parallax, xallarap, and binary lenses.

\acknowledgments
We thank Martin Smith for valuable comments on the manuscript and 
Dale Fields for providing us with the best-fit linear
limb-darkening coefficients for EROS-BLG-2000-5.
Work at OSU was supported by grants AST 02-01266 from the NSF and
NAG 5-10678 from NASA. B.S.G. was supported by a Menzel Fellowship
from the Harvard College Observatory.
C.H. was supported by the Astrophysical Research Center for the
Structure and Evolution of the Cosmos (ARCSEC$"$) of 
Korea Science \& Engineering Foundation
(KOSEF) through Science Research Program (SRC) program.
Partial support to the OGLE project was provided with the NSF grant
AST-0204908 and NASA grant NAG5-12212  to B.~Paczy\'nski and the  Polish
KBN  grant 2P03D02124 to A.\ Udalski. A.U., I.S. and K.\.Z. also
acknowledge support from the grant ``Subsydium  Profesorskie'' of the
Foundation for Polish Science.


\begin{thebibliography}{99}
\frenchspacing

\bibitem[Afonso et al.(2000)]{afonso00}
Afonso, C., et al. 2000, \apj, 532, 340

\bibitem[Albrow et al.(1998)]{planet}
Albrow, M. D., et al. 1998, \apj, 509, 687

\bibitem[Albrow et al.(1999a)]{albrow99}
Albrow, M. D., et al. 1999a, \apj, 522, 1011

\bibitem[Albrow et al.(1999b)]{albrow99b}
Albrow, M. D., et al. 1999b, \apj, 522, 1022

\bibitem[Albrow et al.(2000a)]{albrow00}
Albrow, M. D., et al. 2000a, \apj, 534, 894

\bibitem[Albrow et al.(2000b)]{albrow00b}
Albrow, M. D., et al. 2000b, \apj, 535, 176

\bibitem[Albrow et al.(2001)]{albrow01}
Albrow, M. D., et al. 2001, \apj, 549, 759

\bibitem[Alcock et al.(1993)]{macho93}
Alcock, C., et al. 1993, Nature, 365, 621

\bibitem[Alcock et al.(1995)]{al1}
Alcock, C., et al. 1995, \apj, 454, L125

\bibitem[Alcock et al.(1997)]{alcock97}
Alcock, C., et al. 1997, \apj, 491, 436

\bibitem[Alcock et al.(2000)]{alcock00}
Alcock, C., et al. 2000, \apj, 541, 270 

\bibitem[Alcock et al.(2001)]{alcock01}
Alcock, C., et al. 2001, Nature, 414, 617 

\bibitem[An et al.(2002)]{jin}
An, J. H., et al. 2002, \apj, 572, 521

\bibitem[Aubourg et al.(1993)]{eros93}
Aubourg, E., et al. 1993, Nature, 365, 623

\bibitem[Bessell \& Brett(1988)]{BB88}
Bessell, M. S., \& Brett, J. M. 1988, \pasp, 100, 1134

\bibitem[Bond et al.(2001)]{moa}
Bond, I. A., et al. 2001, \mnras, 327, 868

\bibitem[DePoy et al.(2003)]{depoy}
DePoy, D. L. et al., 2003, SPIE, 4841, 827

\bibitem[Dominik(1998)]{dom}
Dominik, M. 1998, A\&A, 333, 893

\bibitem[Eisenhauer et al.(2003)]{eisen}
Eisenhauer, F., Schoedel, R., Genzel, R., Ott, T., Tecza, M.,
Abutter, R, Eckart, A., \& Alexander, T. 2003, ApJL, submitted
(astro-ph/0306220)

\bibitem[Fields et al.(2003)]{dale}
Fields, D. L., et al. 2003, ApJ, in press.

\bibitem[Gaudi et al.(2002)]{gaudi}
Gaudi, B. S., et al. 2002, \apj, 566, 463

\bibitem[Gould(1992)]{gould92}
Gould, A. 1992, \apj, 392, 442

\bibitem[Gould(1994a)]{gould94}
Gould, A. 1994a, \apj, 421, L71

\bibitem[Gould(1994b)]{gould94b}
Gould, A. 1994b, \apj, 421, L75

\bibitem[Gould(1995)]{gould95}
Gould, A. 1995, \apj, 441, L21

\bibitem[Gould(2000)]{gould00}
Gould, A. 2000, \apj, 535, 928

\bibitem[Gould \& Loeb(1992)]{gouldloeb}
Gould, A. \& Loeb, A., \apj, 396, 104

\bibitem[Gould, Miralda-Escud\'e \& Bahcall(1994)]{gmb}
Gould, A., Miralda-Escud\'e, J., \& Bahcall, J.N. 1994, \apj, 423, L105

\bibitem[Gould \& Welch(1996)]{gouldwelch}
Gould, A. \& Welch, D.L. 1995, \apj, 464, 212

\bibitem[Gradshteyn \& Ryzhik(1965)]{gr}
Gradshteyn, I. S., \& Ryzhik, I. M. 1965, Table of Integrals Series and
Products (London: Academic Press)

\bibitem[Griest \& Safizadeh(1998)]{gs}
Griest, K. \& Safizadeh, N. 1998, \apj, 500, 37

\bibitem[Han \& Gould(1995)]{hangould1}
Han, C., \& Gould, A. 1995, \apj, 447, 53

\bibitem[Han \& Gould(2003)]{hangould2}
Han, C., \& Gould, A. 2003, \apj, 592, 172

\bibitem[Nemiroff \& Wickramasinghe(1994)]{nw}
Nemiroff, R. J., \& Wickramasinghe, W. A. D. T. 1994, \apj, 424, L21

\bibitem[Paczy\'nski(1986)]{bod}
Paczy\'nski, B. 1986, \apj, 304, 1

\bibitem[Paczy\'nski(1996)]{pac98}
Paczy\'nski, B. 1996, Acta Astron., 48, 405

\bibitem[Paczy\'nski \& Stanek(1996)]{pacstan}
Paczy\'nski, B., \& Stanek, K.Z. 1998, \apj, 494, L219

\bibitem[Popowski(2000)]{popow00}
Popowski, P. 2000, \apj, 528, L9

\bibitem[Press et al.(1992)]{numrec}
Press, W. H., Flannery, B. P., Teukolsky, S. A., \& Vetterling, W. T. 1992,
Numerical Recipes (Cambrdge: Cambridge Univ. Press)

\bibitem[Refsdal(1966)]{refsdal66}
Refsdal, S. 1966, \mnras, 134, 315

\bibitem[Rhie et al.(1999)]{mps}
Rhie, S. H., Becker, A. C., Bennett, D. P., Fragile, P. C., Johnson, B. R., 
King, L. J., Peterson, B. A., \& Quinn, J. 1999, \apj, 522, 1037

\bibitem[Salim \& Gould(1999)]{sg}
Salim, S., \& Gould, A. 1999, \apj, 523, 633

\bibitem[Schechter, Mateo \& Saha(1993)]{dophot}
Schechter, P. L., Mateo, M., \& Saha, A. 1993, \pasp, 105, 1342

\bibitem[Smith, Mao \& Paczy\'nski(2003)]{smp}
Smith, M. C., Mao, S., \& Paczynski, B. 2003, \mnras, 339, 925

\bibitem[Smith, Mao, Wo\'zniak(2003)]{smw}
Smith, M. C., Mao, S., \& Wo\'zniak, P. 2003, \apj, 585, L65

\bibitem[Stutz, Popowski \& Gould(1999)]{stutz}
Stutz, A., Popowski, P., \& Gould, A. 1999, \apj, 521, 206

\bibitem[Sumi et al.(2003)]{sumi}
Sumi, T., et al. 2003, \mnras, submitted (astro-ph/0305315)

\bibitem[Udalski et al.(1993)]{ogle93}
Udalski, A., et al. 1993, Acta Astron., 43, 289

\bibitem[Udalski et al.(1994)]{udal}
Udalski, A., Szyma\'nski, M., Ka\l u\.zny, J., Kubiak, M., Mateo, M.,
Krzemi\'nski, W., \& Paczy\'nski, B. 1994, Acta Astron., 44, 227

\bibitem[Udalski et al.(2000)]{ogleII}
Udalski, A., et al. 2000, Acta Astron., 50, 1

\bibitem[Udalski(2003)]{udal03}
Udalski, A. 2003, \apj, 590, 284

\bibitem[Udalski et al.(2002a)]{ogleIII}
Udalski, A., et al. 2002a, Acta Astron., 52, 1

\bibitem[Udalski et al.(2002b)]{ogleIImaps}
Udalski, A., et al. 2002b, Acta Astron., 52, 217

\bibitem[van Belle(1999)]{vanb} van Belle, G. T.\ 1999, \pasp, 111, 1515

\bibitem[Witt \& Mao(1994)]{wm}
Witt, H. J., \& Mao, S. 1994, \apj, 430, 505

\bibitem[Wo\'zniak(2000)]{wozniak}
Wo\'zniak, P. R. 2000, Acta Astron., 50, 421

\end{thebibliography}
\end{document}